\documentclass[12pt]{article}

\usepackage[left=1in,right=1in,top=1in,bottom=1in,letterpaper]{geometry}
\usepackage{sectsty}
\sectionfont{\Large}
\subsectionfont{\large}
\usepackage{caption,setspace}
\usepackage{booktabs}
\usepackage{wrapfig}
\usepackage{subcaption}
\usepackage{graphicx}
\usepackage{multirow,array}
\usepackage{arydshln}
\usepackage{algorithm}
\usepackage{algpseudocode}
\usepackage{comment}
\usepackage[authoryear]{natbib}

\usepackage{hyperref}
\hypersetup{colorlinks,urlcolor=blue,citecolor=black}
\usepackage{authblk}

\usepackage{amsmath, amsthm, amssymb, amsfonts}

\numberwithin{equation}{section}
\usepackage[utf8]{inputenc}

\newcommand{\cT}[0]{\mathcal{T}}
\newcommand{\cI}[0]{\mathcal{I}}
\newcommand{\cS}[0]{\mathcal{S}}

\theoremstyle{remark}
\newtheorem*{remark}{Remark}

\title{Robust domain selection for functional data via\\
interval-wise testing and effect size mapping}
\author[1]{Yeonjoo Park \thanks{Corresponding author: \texttt{yeonjoo.park@utsa.edu}}}
\author[2]{Aiguo Han}
\affil[1]{Department of Statistics and Data Science,
University of Texas at San Antonio}
\affil[2]{Department of Biomedical Engineering and Mechanics, Virginia Tech}
\date{}

\begin{document}

\maketitle

\begin{abstract}
Among inferential problems in functional data analysis, domain selection is one of the practical interests aiming to identify sub-interval(s) of the domain where desired functional features are displayed. Motivated by applications in quantitative ultrasound signal analysis, we propose the robust domain selection method, particularly aiming to discover a subset of the domain presenting distinct behaviors on location parameters among different groups. By extending the interval testing approach, we propose to take into account multiple aspects of functional features simultaneously to detect the practically interpretable domain. To further handle potential outliers and missing segments on collected functional trajectories, we perform interval testing with a test statistic based on functional M-estimators for the inference. In addition, we introduce the effect size heatmap by calculating robustified effect sizes from the lowest to the largest scales over the domain to reflect dynamic functional behaviors among groups so that clinicians get a comprehensive understanding and select practically meaningful sub-interval(s). The performance of the proposed method is demonstrated through simulation studies and an application to motivating quantitative ultrasound measurements.
\end{abstract}

\section{Introduction}
\begin{figure}[!t]
  \centering
  \includegraphics[width=3.1in]{ 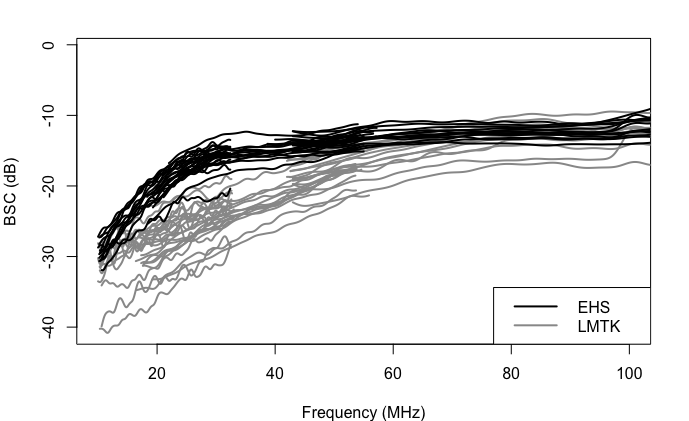}
   \caption{BSC versus frequency from EHS and LMTK tumors}\label{fig:data_intro}
\end{figure}

Statistical methodology for functional data is now a well-developed area with increasingly common continuous monitoring of variables over time or spatial domains from many fields; see, for example, \cite{ramsay2005}, \cite{Ferraty2006}, \cite{horvath2012}, \cite{Morris2015}, and \cite{Wang2016}. In biomedical imaging, quantitative ultrasound (QUS) is one of the ultrasound technologies aiming to facilitate medical diagnosis by extracting quantitative parameters from ultrasound echo signals backscattered from biological tissue. The QUS methodology achieves this goal by providing intrinsic tissue properties that are correlated with tissue physiology, pathologies, or disease processes, such as the amount of fat in the liver \citep{Han2020} and the type and malignancy of a tumor \citep{Han2013, Oelze2016}. Spectral analysis is often used to extract QUS parameters from the ultrasound echo signals. As a result, fundamental QUS parameters such as the attenuation coefficient and backscatter coefficient (BSC) are functional data expressed as a function of frequency \citep{Han2017}. Figure \ref{fig:data_intro} illustrates the BSC data acquired from two different types of implanted mouse tumors, a mouse sarcoma (EHS, ATCC \#CRL-2108) and malignant fibroblast sarcoma (LMTK, ATCC \#CCL-1.3).

In this study, we focus on the domain selection problem aiming to identify interval(s) displaying statistically separable BSC behaviors for different types of tumors and quantify their effect sizes. There have been studies demonstrating the efficacy of QUS measurement as a noninvasive diagnostics tool for tumor screening. Based on experimental studies in \cite{Wirtzfeld2015} demonstrating statistically distinct behaviors of BSC between different types of mammary tumors, \cite{Park2019} developed the probabilistic classifier predicting the tumor type based on BSC trajectories. Later, \cite{Park2022} again confirmed the efficacy of QUS measurements by applying their asymptotically consistent functional Analysis of Variance (fANOVA) type inference procedure. Then, a subsequent interest lies in identifying sub-intervals of the frequency domain displaying such statistically separable features on BSC trajectories. While our motivating data in Figure \ref{fig:data_intro} visually presents more separable group behaviors at low to middle range of frequencies than higher frequencies, we need statistical tools to determine explicit boundaries, which select portions of the domain presenting significant differences among group location parameters. Identifying these sub-interval(s) would increase the accuracy of disease diagnosis and reduce the cost of examination in practice. 

When developing this capability, robustness to outlying trajectories is a particular concern because noninvasive scanning in QUS carries the risk of encountering unexpected contamination, e.g., heterogeneity due to the inclusion of neighboring tissue in the scanned region, as seen from several such measurements in Figure \ref{fig:data_intro}, at the lower frequency ranges. In addition, BSC trajectories were collected by scanning subjects using transducers covering distinct ranges of frequency; thus, trajectories form so-called partially observed functional data structure \citep{Kraus2015,Park2022}, where individual trajectories are collected over specific subintervals within the whole domain. Hence, we need a robust inference tool that can accommodate general functional data structures containing missing segments or irregularities.

Several authors have studied the domain selection problem for fully observed functional data. \cite{Vsevolozhskaya2015} and \cite{Pini2016} considered a multiple local testing approach by taking multiplicity into account for $p$-values calculated from hypotheses testing on equality of means performed on priori-defined finite partitions of domain or from projected basis coefficients by finite-dimensional basis functions, respectively. Although their approaches could control the type-I error under $L_2$ space, they pose a limitation on the potential dependence of inferential conclusions subject to the choice of domain partition or basis functions. To overcome this, \cite{Pini2017} proposed domain selection tools via interval-wise testing under $L_2$ space by adjusting point-wise $p$-values marginally calculated over the continuum domain. However, their extension to the partially sampled data containing potential outlying curves has not been discussed. Also, quantifying degrees of significant separation among different groups has been little explored to date, although such measures are practically useful. 

\begin{figure*}[t!]
  \centering
  \includegraphics[width=5in]{ 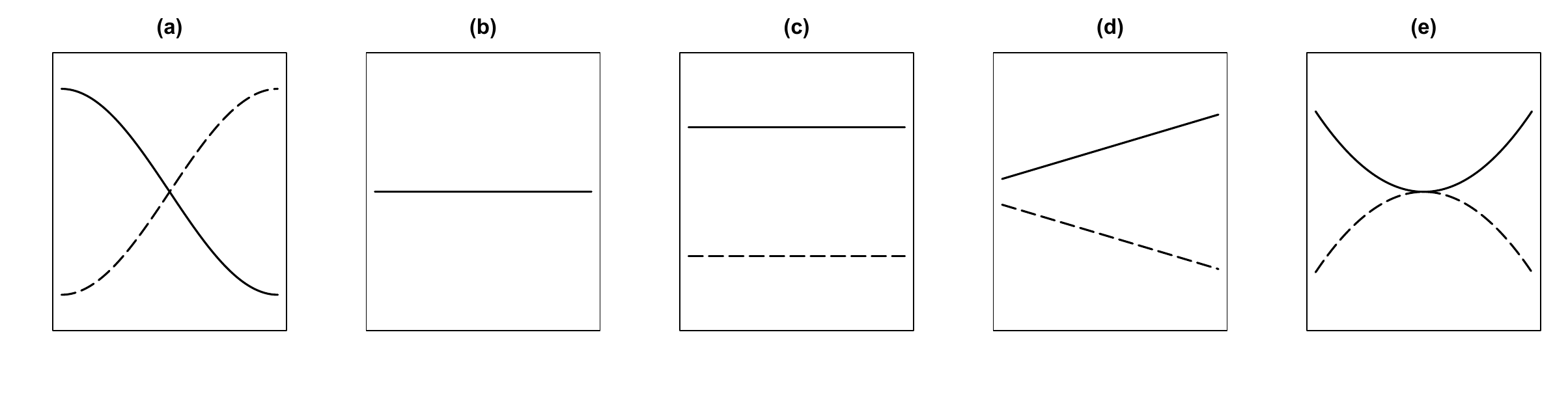}
   \caption{Examples of functional location parameters from two groups}
   \label{Fig:toy_eg_mean}
\end{figure*}

More importantly, we may need to consider other aspects of data, such as first or second derivates of trajectories, simultaneously in hypothesis testing to obtain a clinically interpretable domain. Figure \ref{Fig:toy_eg_mean} illustrates plausible functional behavior scenarios from two group location parameters, one indicated by a straight line and the other marked by a dashed line. While scenarios in (c) and (d) of Figure \ref{Fig:toy_eg_mean} present cases where one group always has significantly larger values than the other group over the domain, Figure \ref{Fig:toy_eg_mean} (a) presents a crossing point in the middle. In the presence of intersecting points, significant functional differences can be masked under the hypothesis of the equality of means, resulting in failure to identify intervals near the cross-over point as a separable domain. To avoid this failure, we should simultaneously consider the first derivatives in testing. In Figure \ref{Fig:toy_eg_mean} (e), two groups behave differently under the second derivative aspect but present marginally close values in the middle. Depending on the goals of the study, discovering the domain featuring distinct functional trends as well as means can be of interest. While \cite{Pini2019} addressed multi-aspect local inference under interval-wise testing, they provide a set of selected domains for corresponding aspects instead of comprehensive identification embracing all considered features simultaneously.

We develop a robust inferential tool that simultaneously considers selected orders of differentiation or other aspects of trajectories to identify unified sub-interval(s) displaying practically interpretable separation. To do this, we robustify the interval-wise inferential procedure \citep{Pini2017} using test statistics calculated from robust M-estimators \citep{Park2022}, which can accommodate partially sampled functional data. To simultaneously reflect inferential interests in multiple features of trajectories, we further take multiplicity correction on robustified adjusted $p$-values. In addition, we introduce the robust effect size map quantifying the degrees of separation between different groups. In practice, significantly distinct differences may not be practically meaningful due to negligible effect sizes. We propose to produce a heatmap providing dynamic effect sizes at each scale by calculating effect sizes from the lowest scale (pointwise manner) to the largest scale (over the whole domain) so that clinicians better understand their functional behaviors.

The rest of the article is organized as follows. In Section 2, we present the robust domain selection inferential tools applicable to partially observed functional data based on an interval-wise testing approach. This section also introduces the robust effect size map with examples of its interpretation. We conduct simulations to examine its selection performance in Section 3. We apply the proposed method to our motivating QUS data in Section 4 and conclude the article with discussions in Section 5.

%
%
\section{Methodology}
\label{sec2:method}
We embed the testing problem in the Sobolev space denoted as $H^L(\mathcal{T})$, consisting of all real-valued square-integrable functions on the domain $\mathcal{T}$ with square-integrable derivatives up to order $L$, where $\mathcal{T} = (a, b) \subset \mathbb{R}$. Let $(\Omega, \mathcal{F}, P)$ be a probability space on $H^L(\mathcal{T})$ and assume that $X_{g1}, \ldots X_{g n_g}$, $g=1, \ldots, k$, denote $k$ groups of random samples drawn from random function $ \mathcal{X}_g$, mapping from $\Omega$ to $H^L(\mathcal{T})$. Let $\mu_g$ denote the location parameter of $g$th group, for example, mean, median, or quantile functions. For mean, $\mu_g = \mathbb{E}[ \mathcal{X}_g]$. For median, $\mu_g = \arg\min_{h \in H^L(\mathcal{T})} = \mathbb{E}[ \| \mathcal{X}_g -h \|_{H^L} - \| \mathcal{X}_g \|_{H^L}]$, where $\| f \|_{H^L} = \sum_{\ell=0}^L \int_{\mathcal{T}} |f^{(\ell)}(t)|~dt $ with $f^{(\ell)}(t)$ denoting the $\ell$th derivative of function. This definition extends the concept of the geometric median under a Hilbert space, as proposed by \citep{Godichon-Baggioni2016}. It is often interesting to test the equality of the $k$ group location parameters in the usual $L^2$ sense, that is, 

\begin{equation}\label{eqn:fanova.V2}
 H_0: \mu_1 = \cdots=\mu_k  \Leftrightarrow \int_\mathcal{T} \{ \mu_g(t) - \mu_{g'} (t)\}^2 dt = 0 \textrm{~for~} \forall g \neq g',
\end{equation}
against the alternative that at least two location functions are not equal over certain subset(s) of $\cT$. The functional Analysis of Variance (fANOVA) is a special case where testing the equality of functional means is of interest.

We note that \eqref{eqn:fanova.V2} focuses on detecting the evidence of significant differences among mean or location parameter functions for any $t \in \cT$, and there have been various $F$-type test statistics using functional mean or M-estimators \citep{faraway1997regression,Zhang2014,  Park2022}. Under a similar setting, \cite{Pini2017} proposed the interval-wise testing aiming to select portions of domains displaying significant differences, referred to as domain selection. It provides a more refined conclusion beyond the rejection of the global fANOVA null hypotheses by discovering portions of the domain that exhibit significant differences in group location parameters. 
While multiple correction methods in hypothesis testing, in general, aim to control the {\textit{family-wise error rate}} (FWER) among a finite number of hypotheses, the domain selection problem in functional data analysis involves a continuous infinity of univariate test from each $t$ over the domain of interest. Owing to this fundamental difference, \cite{Pini2017} introduced the adjusted $p$-value function and proposed to select intervals of the domain by thresholding it at level $\alpha$ to {\textit{interval-wise}} control the probability of type-I error and achieve {\textit{interval-wise}} consistency. The following section reviews the interval-wise testing and the meaning of {\textit{interval-wise}} error control.

\subsection{Review of interval-wise testing for functional data} \label{ssec:review_iwt}
Let $\mathcal{I} \subseteq \cT$ be an generic interval of the form $(t_1, t_2)$, where $a \leq t_1 < t_2 \leq b$, or complementary of the interval $\cT~\textrm{\textbackslash}(t_1, t_2)$. We now consider
\begin{equation}\label{eqn:iwt}
H_0^\mathcal{I}: \mu_1^\mathcal{I} = \cdots = \mu_k^\mathcal{I}, 
\end{equation}

where $\mu_g^\mathcal{I}$ denotes the restriction of $\mu_g$ over $\mathcal{I}$ and let $p^\mathcal{I}$ denote the $p$-value calculated from this test. Here, the choice of test statistics can be flexible depending on data distribution, including parametric \citep{faraway1997regression, Zhang2014}, nonparametric \citep{Pini2017}, or robustified testing, where robust statistic will be introduced in the next section. \cite{Pini2017} then proposed unadjusted and adjusted $p$-value functions, denoted as $p(t)$ and $\tilde p(t)$, respectively, based on $p^\cI$, and their formal definitions are as below. 
\begin{equation} \label{eqn:unadj_pval}
p(t) = \limsup\limits_{\cI \rightarrow t} p^\cI, \quad \quad  
\tilde p(t) = \sup\limits_{ \cI \ni t } p^\cI, \quad \quad \forall t \in \cT,  
\end{equation}
where $\cI \rightarrow t$ indicates that both extremes of interval $\cI$ converge to $t$. 

Two different definitions of $p$-value functions present different inferential properties. While detailed asymptotic type-I error and power derivations are specified in Theorems A.1 - A.4 of \cite{Pini2017}, adjusted $p$-value function $\tilde p(t)$ provides control of interval-wise error rate; that is, for a given level $\alpha \in (0,1)$, any $\cI \subseteq \cT$ where $H_0^\cI$ is true, $P(\forall t \in \cI, \tilde p(t) \leq \alpha) \leq \alpha$. It heuristically implies that if a thresholding at level $\alpha$ is applied to $\tilde{p}(t)$, for each interval of the domain where $H_0$ is true, the probability that $H_0$ is rejected on the entire interval is less than or equal to $\alpha$.  In addition, the interval-wise consistency holds, implying that for each interval $\cI$ including $t$ displaying significant differences among $\mu_g(t)$, the probability of being an entirely selected interval converges to one as the sample size increases. We refer readers to Remarks 2.1 - 2.5 of \cite{Pini2017} for a comprehensive explanation. 

\subsection{Robust domain selection using functional M-estimators}
\label{ssec:domain_selection}

To detect practically interpretable differences, raw trajectories as well as up to their desired order of derivatives should be simultaneously considered. Let $\ell \in \{0,1, \ldots, L\}$ indicate the order of derivatives we take into account. With trajectories potentially containing heavy-tail features, we consider the simultaneous interval-wise testing under
\begin{equation}\label{multiple_null}
H_{0}^\cI: \theta_{1, D^\ell}^\cI = \cdots = \theta_{k, D^\ell}^\cI \quad\textrm{for}\quad \ell = 0,\ldots, L,
\end{equation}
where $\theta_{g, D^\ell}$ denotes the robust location parameter for group $g$ from $\ell$th derivatives of functions. The rejection of \eqref{multiple_null} indicates at least two unequal $\theta_{g, D^\ell}^\cI$ and $\theta_{g', D^\ell}^\cI$ from at least one order of $\ell$. For implementation, we propose calculating adjusted $p$-value functions using robust test statistics based on functional M-estimators of each order of $\ell$ and applying a multiple testing correction. 

We consider functional data from a partial sampling scheme as in motivating data, where functional trajectories $X_{g1}(t), \ldots, X_{g n_g}(t)$ are collected over individual-specific subsets $\cS_{g1}, \ldots, \cS_{g n_g} \subseteq \cT$, where $\cup_{g,i} \cS_{gi} = \cT$. To formulate this missing framework, we introduce an observation indicator function $\delta_{gi}(t)$, $t \in \cT$, that $\delta_{gi}(t) = 1$ if $t \in \mathcal{S}_{gi}$, and  $\delta_{gi}(t)=0$, otherwise. Below are the listed assumptions for this indicator function.

\begin{itemize}
    \item[(A1)] The observation indicators $\delta_{gi}$, $g=1, \ldots,k$, $i=1, \ldots, n_g$ are i.i.d. with $\mathbb{E} [\delta_{gi}(t)] = b(t)$, where $\inf_{t \in \mathcal{T}} b(t) >0$, and $\mathbb{E} [ \delta_{gi}(t) \delta_{gi}(t')] = v(t,t') < \infty$.
    \item[(A2)] $X_{gi}(t)$ and $\delta_{gi}(t)$ are independent.
\end{itemize}
Condition (A1) implies that the full domain of the interest is covered by a sufficient portion of the data from sufficiently large sample sizes. The bounded covariance function condition guarantees the finite covariance function of the functional M-estimator, which shall be presented in the following paragraph. Condition (A2) indicates missing completely at random.

In practice, partially observed trajectories are evaluated over a set of discrete grid points in $\mathcal{S}_{gi}$, that is, $ D_{gi} =\{ t_{gij} \in \cS_{gi}$, $j=1, \ldots, m_{gi}\}$. For simplicity, we focus on the regular grid design $ D_{gi} \subseteq D$, where $ D =\{ t_j \in \cT, j=1, \ldots, m\}$, implying partially observed trajectories evaluated at a regular grid design. But evaluation over irregular grid points does not affect our methodological developments. We then obtain the functional M-estimator over $\cT$ for group $g$ under $\ell=0$, denoted as $\hat\theta_{g, D^0}$, as a function $h \in H^L(\mathcal{T})$ that minimizes 
\begin{equation}\label{eqn:M_est_rho_ps}
 \sum_{i=1}^{n_g}  \sum_{j=1}^{m} \frac{1}{\sum_{j} \delta_{gi}(t_{j})} \delta_{gi}(t_{j}) \rho \{  X_{gi}(t_{j}) - h(t_{j})  \} + \lambda \int_\cT|h^{(L)}(t)|^2 dt,
\end{equation}
where $\lambda$ governs the smoothness of $h$ and the robust loss function $\rho(\cdot)$ controls the degree of robustness by weakening the influence of atypical values. Examples include Huber or bisquare loss functions. Let $f^{(\ell)}(t)$ denotes the $\ell$th derivatives of $f(t)$, then the functional M-estimator under $\ell > 0$ can be obtained using a similar objective function $\eqref{eqn:M_est_rho_ps}$ by replacing $X_{gi}(t_j)$ and $h^{(L)}(t)$ with $X_{gi}^{(\ell)}(t_j)$ and $h^{(L-\ell)}(t)$, respectively, under $h \in H^{L-\ell}(\mathcal{T})$. \cite{Kalogridis2022} presents technical conditions for the existence of solutions and rates of convergence under fully observed trajectories, along with practical computation of solutions under the B-spline space. See Remark \ref{rem-computation} for computational details to find a solution.

We then calculate the interval-wise test statistic $T^{\cI}_{D^\ell}$ by applying a robust functional ANOVA as, 
 \begin{equation}\label{eqn:robust_fANOVA}
    T^\cI_{D^\ell} =  \frac{1}{|\cI|} \sum_{g=1}^k \int_{\cI} \alpha_g \{\hat \theta_{g,{D^\ell}}(t) - \bar \theta._{,D^\ell} (t)\}^2 dt,
 \end{equation}
where $\alpha_g = n_g/n$ under $n=\sum_{i=1}^k n_g$, $\bar \theta._{,D^\ell} (t) = \sum_{g=1}^{k} \alpha_g \hat\theta_{g,D^\ell} $ denotes the weighted grand mean, and $|\cI|$ is the length of the interval $\cI$. The corresponding $p^\cI_{D^\ell}$ under each $\ell$ in \eqref{multiple_null} is calculated from a bootstrap procedure.

Now, let $ \tilde{p}_{D^\ell}(t)$ denote the calculated adjusted $p$-value functions based on robust test statistic $T^\cI_{D^\ell}$, for $\ell = 0, \ldots, L$. Since multiple claims are considered simultaneously in the inference process, further correction is needed depending on the selected $L$. Among multiple testing correction methods, we aim to control Family-Wise Error Rate (FWER) and apply corrected $\alpha^*$ to each $\tilde{p}_{D^\ell}(t)$. The corrected $\alpha^*$ can be from the Bonferroni, Holm-Bonferroni \citep{Holm1979}, or Hochberg’s correction method \citep{Hochberg1988}. Then, under $ \tilde{p}_{D^\ell}(t)$ with control of type-I error and adjusted $\alpha^*$, interval-wise type-I error is preserved; that is, for a given level $\alpha \in (0,1)$, any $\cI \subseteq \cT$ where $H_0^\cI$ is true, $P( \textrm{at least one}~ \ell \in \{0, \ldots, L \}, \forall t \in \cI ~~ \tilde p_{D^\ell}(t) \leq \alpha^*) \leq \alpha$. With the corrected $\alpha^*$, we then subsequently identify the sub-interval(s) of the domain displaying significant group differences on robust location parameters from at least one order $\ell$ by
\begin{equation}\label{domain-select}
C = \cup_{\ell=0}^L C_{\ell},~~ \textrm{where}~~ C_\ell = \{ t \in \cT: \tilde{p}_{D^\ell}(t) \leq \alpha^* \} . 
\end{equation}

\begin{remark}\label{rem-presmooth}
In practical domain selection, trajectories should be pre-smoothed to ensure they reside in the assumed $\mathcal{H}^L(\mathcal{T})$ and to obtain the corresponding derivative trajectories. In this data preparation stage, kernel-weighted local polynomial smoothing can be applied, with the bandwidth selected via cross-validation (CV). The function \texttt{locpoly()} in the R package \texttt{KernSmooth} is then used to compute derivatives up to order $L$. Both the simulation studies and the real-data analysis in this paper adopt this kernel-based pre-smoothing approach. Alternatively, individual trajectories may be approximated using B-spline basis functions through the R package \texttt{fda}, with derivatives up to the desired order $L$ computed using \texttt{deriv.fd()}.

\end{remark}

\begin{remark}\label{rem-computation}
Based on the pre-smoothed data, we find a solution minimizing $\eqref{eqn:M_est_rho_ps}$ under a nonparametric regression approach. Specifically, we represent $h$ as a penalized polynomial smoothing spline with a desired order. Let $h(t) = \sum_{k=1}^d c_k \phi_k(t)$, where $\phi_k(t)$, $k = 1, \ldots, d$, denote spline basis functions constructed from a sequence of knots selected via cross-validation (CV), and where the polynomial order depends on $L$. The roughness penalty parameter $\lambda$ is selected using generalized cross-validation (GCV). The resulting estimator is computed through an iterative reweighted least-squares (IRWL) algorithm under the robust loss function $\rho(\cdot)$, by solving a weighted penalized least-squares problem at each iteration until convergence \citep{maronna2019robust}. The detailed algorithm with the Huber loss function is provided in Section S1 of the Supplementary Material, and the corresponding R implementation is available on GitHub \href{https://github.com/ypark0917/Robust-domain-selection-for-functional-data-via-interval-wise-testing-and-effect-size-mapping}{(link)}.

\end{remark}

\begin{remark}
The determination of $L$ depends on the goal of the study. In our motivating application with BSC data, $L=1$ is considered because the main research interest is to discover the domain displaying distinct tendencies on raw and first derivative trajectories. However, if comparing the accelerating rate of change is of interest, for example, for growth curves, $L$ might be set as 2.
\end{remark}

\subsection{Robust effect size heatmap}
\begin{figure*}[t!]
  \centering
  \includegraphics[width=6.8in]{ 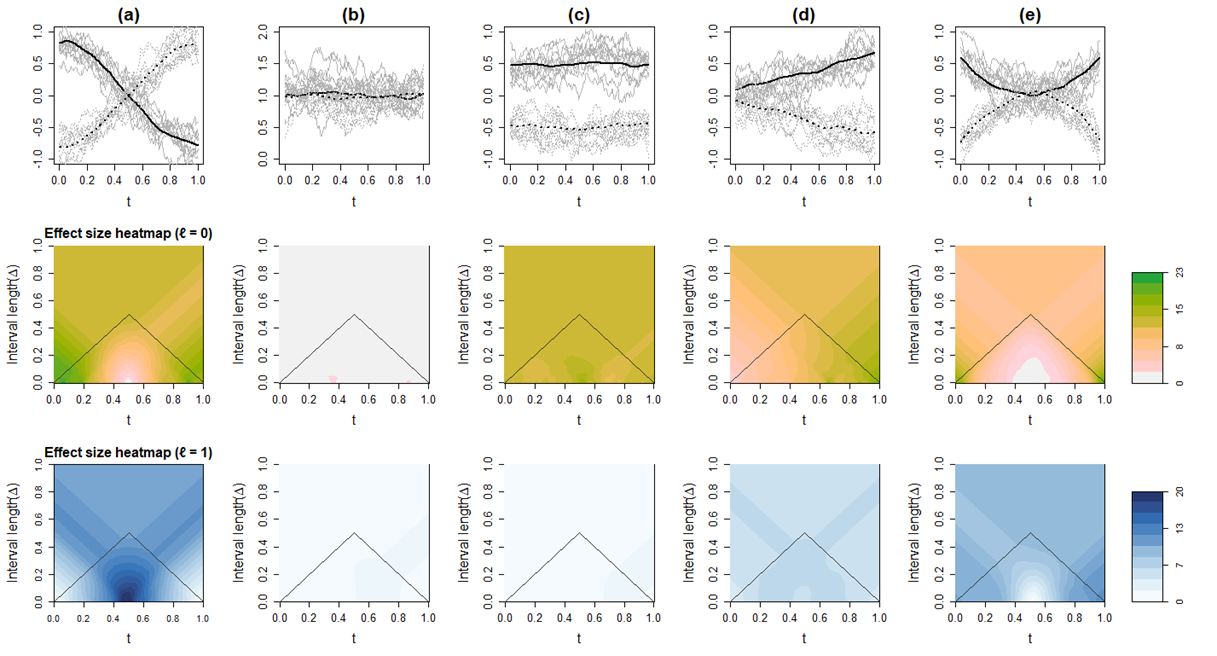}
   \caption{Illustrations of effect size heatmaps for $\ell=0$ (second row) and  $\ell=0$ (third row) from various scenarios, calculated under \eqref{eqn:es} from fine to coarse scales of $\Delta$. Effect size values within the triangle region are calculated based on the symmetric choice of lower and upper extremes, where the corresponding $\Delta$ in the y-axis exactly matches the length of the interval integrated to obtain $G^2_{fSNR}(t; \Delta)$. Values outside the triangle region are derived from asymmetric lower and upper extremes, implying the actual integrated interval length is shorter than $\Delta$.}\label{Fig:effectsize_example}
\end{figure*}

Along with the selection of intervals, quantifying the degree of distinction is crucial to identifying the subset of the domain featuring clinically meaningful group separation. We propose to examine the effect size heatmap at each scale, where the effect sizes from the lowest scale (i.e., pointwise size) are displayed at the bottom, and the result from the largest scale (i.e., average effect size over the entire domain) is shown at the top. Figure \ref{Fig:effectsize_example} illustrates the effect size heatmap under $\ell=0$ from scenarios considered in Figure \ref{Fig:toy_eg_mean}. The key is to present its effect size in all scales so that local cross-over points do not mislead practitioners in the interpretation. 

For notational simplicity, let $\hat \theta_g(t)$ denote $\hat\theta_{g, D^0}(t)$ and by extending the robust effect size index \citep{Vandekar2020} to functional context, we define the robustified functional Signal-to-Noise Ratio ($fSNR$) for raw trajectories as, 
\begin{equation} \label{eqn:fSNR}
fSNR^2(t) = \frac{\sum_{g=1}^k n_g \{\hat\theta_g(t) - \bar \theta.(t)\}^2}{\hat\xi_n^2(t)}~, \quad t \in \cT,
\end{equation}
where $\hat\xi_n^2(t) := \hat\xi_n(t,t)$ denotes the estimated variance of $\hat\theta_g(t)$ at $t$. The subscript $n$ instead of $n_g$ is due to variance estimation via data pooling across $g$ under the homogeneous covariance structure assumption across $g$. 

As demonstrated in \cite{Vandekar2020}, this robust index \eqref{eqn:fSNR} yields several classical effect size indices, such as Cohen’s $d$ or $R^2$, when the models are correctly specified. For practical implementation, we estimate $\xi(s,t)$ using the bootstrap samples.
Then effect size heatmap is generated under different scales as below. For each $t \in \cT$ and the scale $0<\Delta < |\cT|$, we calculate the aggregated effect size,
\begin{equation}\label{eqn:es}
G^2_{fSNR}(t; \Delta) = |u_{(t,\Delta)} - l_{(t,\Delta)}|^{-1} \int_{l_{(t,\Delta)}}^{u_{(t,\Delta)}} fSNR^2(t') dt' 
\end{equation}
where $l_{(t,\Delta)}= \max\{a, t-(\Delta/2) \}$ and $u_{(t,\Delta)}=\min\{ b, t+ (\Delta/2) \} $. We then display the results from fine to coarse scales up to the integrated $fSNR(t)$ over the whole domain $\cT$, as in Figure \ref{Fig:effectsize_example}.

The effect size heatmap helps understand the behavior of group differences. As in Figure \ref{Fig:effectsize_example} (c), when the map shows similar effect sizes at all scales, it suggests no cross-overs among group locations parameters, achieving similar degrees of separation over $\cT$. Figure \ref{Fig:effectsize_example} (a) displays a small or even zero effect size at the middle regions at the fine-scale but shows gradually increasing effects as the interval widens. It implies the cross-over group parameter functions with neighbors of such $t$ reveal distinct group behaviors afterward; thus, these regions are expected to differentiate groups. However, in practice, interpretation should be based on a comprehensive investigation, such as visualizing the functional group M-estimators along with their estimated marginal variances, to assess the factors contributing to a small effect size, whether due to cross-points or lack of distinction. Next, the small effect size at the fine level, left part of the domain in Figure \ref{Fig:effectsize_example} (d) shows similar small sizes across nearby neighbors and gradual changes in one direction, indicating practically non-separable features. Lastly, we note that the effect size heatmap can also be generated with any order of derivatives with the corresponding estimation of $fSNR(t)$, if it is of interest. The implementation R code can be found in Supplementary Material.

%
%
%
\section{Simulations} \label{sec:sumulation}
We conduct simulation studies to evaluate the domain selection performance of our proposed method. To do this, we generate $n$ random trajectories from two groups, respectively, using  $\mu_1(t)$ and $\mu_2(t)$, two distinct functional location parameters. To mimic the motivating BSC application, $\mu_1(t)$ is specifically generated by computing the theoretical BSC as a function of frequency for acoustic backscattering from a fluid sphere of diameter 10 $\mu$m embedded in a uniform fluid background \citep{Anderson1950}, a theoretical model commonly used in the QUS literature \citep{Han2023}. And $\mu_2(t)$ is generated similarly for a fluid sphere of diameter 11 $\mu$m. We then artificially align  two group parameters over $[0,c_1]$ by forcing $\mu_2(t) = \mu_1(t)$ for $t \in [0, c_1]$ and $c_1=0.34$, so that two groups are separable over one sub-interval $(c_1,1]$.  The additional case with two disjoint sub-intervals exhibiting separable group behavior is also considered, while experiment details and results are deferred to the Supplementary Material.

\begin{figure*}[t!]
    \centering
    \includegraphics[width=6.5in, height = 1.8in]{ 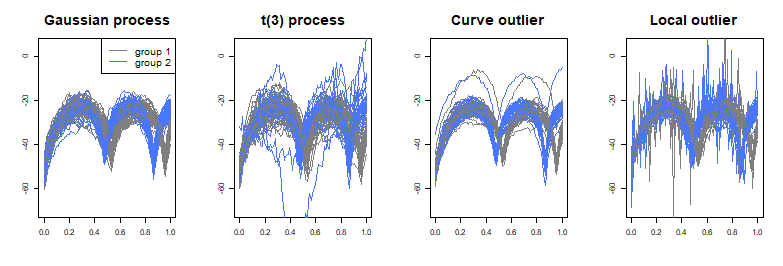}
    \includegraphics[width=6.5in]{ 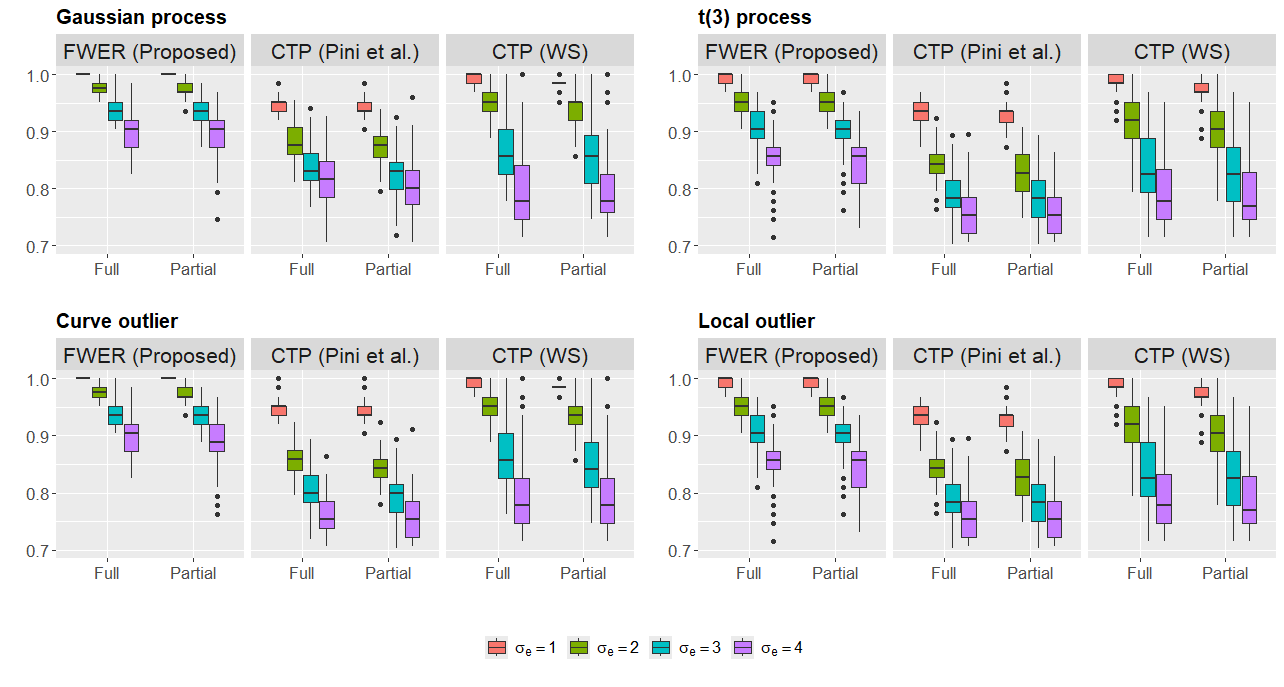}
    \caption{Illustration of fully observed trajectories from the scenarios of Gaussian process, $t_3$ process, curve outlier, and local outlier under the noise level $\sigma_e=3$ (top row). Boxplots of sensitivity from the proposed method and the comparison method, two variations of CTP-based method, under fully and partially observed trajectories and noise levels $\sigma_e = 1, 2, 3, 4$, when sample size $n = 50$.} 
    \label{Fig:simulation_rep}
\end{figure*}

Next, based on $X_{gi}(t) = \mu_g(t) + e_{gi}(t)$, $ t \in [0,1]$, for $g=1,2$, $i=1, \ldots, n$, we generate $e_{gi}(t)$ from the mean-zero process under four scenarios: (i) Gaussian error process, i.e., $e_{gi} \sim GP(0, \gamma_e)$, (ii) $t_3$ error process, i.e., $e_{gi} \sim t_3(0, \gamma_e)$, (iii) curve outlier where a subset of trajectories display outlying behaviors over the entire domain, and (iv) local outlier where a subset of trajectories are contaminated by local spikes. Functional trajectories are evaluated on a regular grid of 100 points in $[0, 1]$, and to be specific, for scenarios (i) and (ii), we employ the exponential scatter function ${\gamma_{e}}(d) = \sigma_e^2 \exp(-d/\phi)$, where $d=d(t, t^{'})$ denotes the distance between two points and $\phi$ represents the range parameter determining the spatial dependence within a curve. Here, we note that our error generation is under stochastic perspective with exponential autocovariance function rather than the Hilbert space perspective, described in the section of methodology, because outlying curve generation is more straightforward under stochastic view. While smaller (larger) $\phi$ indicates weaker (stronger) dependence, the simulations use $\phi=0.2$, representing the moderate dependence over $t$. We note that the variation of the $t_3$ process is three times larger than that of the Gaussian process as a family of elliptical processes \citep{Park2023}. Then, curve outliers are generated by applying abnormal shifts over the entire domain to a subset of trajectories generated from the Gaussian error process. In detail, randomly selected 5\% of total trajectories from scenario (i) are contaminated by adding random shifts over the entire domain, where such shift is independently sampled from $t$ distribution with 3 degrees of freedom and the variance of $3\cdot \sigma_e^2$. Next, the local outlier is again generated based on trajectories from the Gaussian error process by contaminating them with heavy-tailed local noise. By randomly choosing 5\% of grids from a pool of grids from trajectories of scenario (i), we add additional noise, independently generated from $t$ distribution with 3 degrees of freedom with the variance $2 \cdot \sigma_e^2$. Four noise levels $\sigma_e = 1,2,3,4$ are considered to reflect different effect sizes of the signal and we set $n=50$ or 100. The top row of Figure \ref{Fig:simulation_rep} illustrates generated trajectories under four scenarios under $n=50$ and $\sigma_e = 3$, and we observe that scenario (ii) represents the functional outliers displaying atypical behaviors over the subset of the domain.

We then further consider two sampling frameworks: fully observed or partially observed responses. For the partial sampling structure, we generate an independent random missing interval $M_{ij}$ for each trajectory on which functional values are removed. An indicator variable $B_i \overset{i.i.d,}{\sim} Bernoulli(0.5)$ is generated, and if $B_i=0$, we set $M_{ij}$ as a null set, meaning no missing segments in the $i$th trajectory. If $B_i=1$, $M_{ij}$ is generated from $M_{ij}=[C_{ij} - E_{ij}, C_{ij}+E_{ij}] \cap [0, 1]$ by following \cite{Kraus2015}, where $C_{ij}=d U_{ij,1}$ and $E_{ij}=f U_{ij,2}$ with i.i.d. uniformly distributed $U_{ij,1}, U_{ij,2}$ on $[0, 1]$, and constant parameters $d$, $f$, set as $d =1.2$ and $f=0.3$. This missing framework results in 25.6\% of each trajectory being removed by this missing interval $M_{ij}$ on average for observations with non-null $M_{ij}$. Putting all this together, we consider the combinations of four scenarios and two sampling schemes under four noise levels and $n=50$ or 100. In this study, we consider the first-order derivative of trajectories along with the raw data simultaneously to detect the distinct behaviors.

As mentioned in Remark \ref{rem-presmooth}, trajectories should be pre-smoothed using either kernel- or spline-based methods. In the simulation study, kernel-based smoothing is adopted by choosing the bandwidth based on CV errors.
Next, we apply the proposed method by calculating the interval-wise robustified test statistic \eqref{eqn:robust_fANOVA} based on functional M-estimators computed for each group under the robust tuning parameter $\delta=1$ for the Huber loss function $\rho$ in \eqref{eqn:M_est_rho_ps}. Considering that the optimal choice of $\delta$ for Gaussian data is known as $\delta=1.35$ and $\delta=0.8$ empirically gives the estimates close to the median, we choose the intermediate tuning parameter. The resulting conclusions are not sensitive to the choice of $\delta$ unless they are too small or large. Also, we apply four correction methods for simultaneous feature consideration discussed in the previous section, from the Bonferroni to other FWER methods, with results displayed under the most conservative Bonferroni method in the manuscript. However, owing to the small $L=1$, we observe almost the same inferential conclusions for all considered multiple testing correction methods.

\cite{Pini2019} similarly considered the multiple orders of derivatives of trajectories in the domain selection problem by combining the IWT approach and the Close Testing Procedure (CTP), a multiple-testing correction technique. The major distinction between our proposed and multi-aspect IWT lies in the choice of multiplicity correction method, where CTP involves a single test considering all orders of derivatives jointly. While methodological details of multi-aspect IWT can be found in \cite{Pini2019}, we perform it using the test statistics based on the functional M-estimator for a fair comparison to our method. We specifically consider two variations of the comparison methods, labeled `CTP (Pini et al.)' and `CTP (WS)'. The former computes an aggregated test statistic as the summation of test statistics, $T^\cI_{D^\ell}$, while the latter rescales test statistics from each $\ell$ before aggregation.

\begin{table*}[t!]
\begin{center}
\caption{The average false rejection rate (FRR) and the probability of including at least one false rejection on $\mathcal{A}^c$ over 100 repetitions calculated from four scenarios of functional behaviors for fully and partially observed trajectories under $n=50$ and $\sigma_e = 4.$}
\vspace{3mm}
\begin{tabular}{ccccc}
    \hline
       & Sampling structure & FRR & $P$(at least one false rejection) \\
    \hline
    \multirow{2}{*}{Gaussian process}& Full & 0.01& 0.05 \\
     & Partial & 0.004 & 0.02 \\
              \hline
    \multirow{2}{*}{$t_3$ process}& Full& 0.004& 0.03\\
     & Partial& 0.001 & 0.01 \\
         \hline
    \multirow{2}{*}{Curve outlier}& Full& 0.01 & 0.06 \\
     & Partial& 0.005 & 0.04 \\
    \hline
        \multirow{2}{*}{Local outlier}& Full& 0.01 & 0.03 \\
     & Partial& 0.01 & 0.02 \\
    \hline
    \label{tab:FDR}
\end{tabular}
\end{center}
\end{table*}
{\color{black} We then evaluate domain selection performances under different scenarios and settings through 100 repetitions. Let $\mathcal{A} = \{t_l: \mu_1(t_l) \neq \mu_2(t_l),~l=1, \ldots, 100 \}$ be the subset of $\{t_1, \ldots, t_{100} \} \in [0,1]$ containing grids displaying distinct separation between two groups. And let $\hat{ \mathcal{A}}$ denote the estimated separable domain from the proposed or comparison method. We then examine three measures: (i) sensitivity as $ \#\{t_l: t_l \in \mathcal{A}~\mbox{and}~ t_l \in \hat{\mathcal{A}} \}/ \# \mathcal{A}$, where a sensitivity of 1 indicates perfect selection of domain displaying separable group behaviors, (ii) false rejection rate (FRR) as $\#\{t_l: t_l \in \mathcal{A}^c~\mbox{and}~ t_l \in \hat{\mathcal{A}} \}/\# \mathcal{\hat A}$, the lower the better, and (iii) the probability of at least one false rejection among $\mathcal{A}^c$. While the first two measures are calculated for each simulated data and the averages among them are presented as the result, the last measure is calculated based on the number of simulation sets having a non-zero false rejection rate among repetitions. }

{\color{black} The middle and bottom rows in Figure \ref{Fig:simulation_rep} display sensitivity under $n=50$ from the proposed and comparison methods, and these rates decrease as the noise level increases, as expected. Although missing segments or outlying noises mask true distinctions, the performance of our method seems favorable. We also observe similar performances under full and partial structures, implying the practical utility of our methods even under missing segments. Except for the challenging scenario with s large noise level $\sigma_e=4$ and partial sampling structure, overall sensitivities show desirable performance with rates mostly above 90\% regardless of the categories of outliers. In terms of two comparisons based on CTP, their sensitivities show mostly lower than those from our proposed method, and we especially observe a relatively bad performance in identifying the separable domain that features a small magnitude of mean difference between the two groups but almost similar behaviors in the first order of derivative. We presume that the test statistics jointly considering all orders may reduce the chance of rejection when only one aspect has a significant difference with a small effect size, but no significant difference for other aspects. Adopting an alternative aggregation through the rescaled test statistic did not bring significant improvement. A similar phenomenon is observed under $n=100$ as well, although it is not presented here. The boxplots of sensitivity of our proposed method under $n=100$ showing almost perfect identification attained under $\sigma_e=1$ or 2, even under a partial sampling structure, are presented in the Supplementary Material, 
}

{\color{black} Table \ref{tab:FDR} displays FRR and the probability of including at least one false rejection from our method under $n=50$ and $\sigma_e = 4$ from fully and partially observed trajectories. This combination indeed represents the most challenging scenario. We first observe that FRR is relatively low in all cases, implying that our method makes less false rejection for the domain featuring non-separable group behaviors. Also, the probability of including at least one false rejection on $\mathcal{A}^c$ over repetitions ranges from 0.01 to 0.03, implying that, for most of the simulation sets, our method could successfully find the domain with the non-separable group behaviors. It indicates that making the correct rejection on $\mathcal{A}$, especially for the domain presenting a small difference between groups, is the crucial part of the domain selection problem. }
\begin{figure*}[t!]
  \centering
  \includegraphics[width=7in]{ 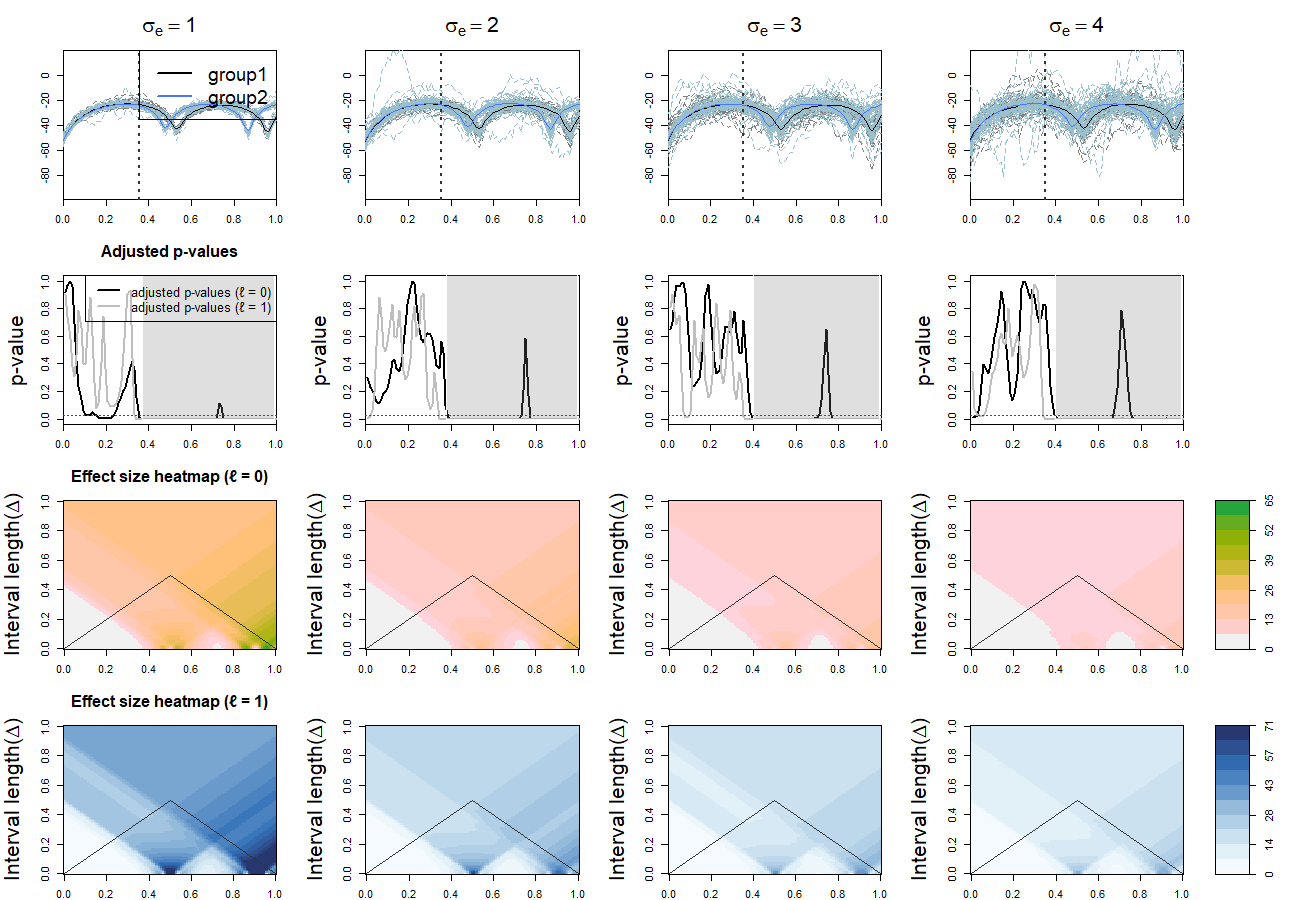}
   \caption{{(Top row) Simulated partially observed trajectories under $t_3$ error processes and $n=100$, where $\sigma_e=1,\ldots,4$, respectively, where black and blue bold lines indicate true group location parameters featuring distinction over $t \in (0.34,1]$ with a dotted vertical line locating at $t=0.34$; (Second rows) Adjusted p-value functions $\tilde{p}_{D^0}(t)$ and $\tilde{p}_{D^1}(t)$ for testing on equality of two group parameters from trajectories of raw and the first-order derivatives. The gray regions illustrate the selected intervals under the proposed method; (Third and fourth rows) The robust effect size heatmaps under $\ell=0$ and 1, respectively. Effect size values within the triangle region are calculated based on the symmetric choice of lower and upper extremes (integrated interval length equals to $\Delta$), while values outside the triangle region are derived from asymmetric lower and upper extremes (integrated interval length shorter than $\Delta$).}}\label{Fig:simulation}
\end{figure*}

Lastly, we illustrate the selection result and estimated effect size heatmap from the proposed method under the partially sampled $t_3$ error process scenario. The top four plots of Figure \ref{Fig:simulation} illustrate generated trajectories from $\sigma_e=1, \ldots,4$, respectively, under $n=100$, where bold lines represent true group location functions featuring separable distinctions over $(0.34, 1]$. The vertical dotted line is located at $t=0.34$ for reference. Then, plots in the second row of Figure \ref{Fig:simulation} display adjusted $p$-value functions, $\tilde{p}_{D^0}(t)$ and $\tilde{p}_{D^1}(t)$, from hypothesis testing on equality of two group parameters using raw and first-order derivative trajectories, respectively. 
The gray-highlighted region displays the selected domains from the proposed methods. Even with striking outliers under the partial sampling structure, our method could detect domains with almost perfect performance under $\sigma_e=1$ or 2. The performances under $\sigma_e=3$ or 4 are also desirable by identifying most of separable domains except for a small portion displaying practically negligible differences. Our method successfully contains cross-over points in the detected region by taking into account the first-order derivatives. The heatmaps in the last two rows of Figure \ref{Fig:simulation} show dynamic effect size information under $\ell = 0,1$. Under $\ell=0$, we observe overall zero or a small degree of group distinction for $t$ below 0.4, but practically separable behaviors shown above 0.4, especially the strongest signals observed from $t$ around 0.8 to 1 from all cases. Among four noise levels, the most separable distinctions were observed from $\sigma_e=1$. We observe similar phenomenons with strong signals near $t=0.5$ and 0.9 under $\ell=1.$

%
%
%
\section{Application to Quantitative Ultrasound signal analysis}
\begin{figure*}[t!]
  \centering
  \includegraphics[width=5.8in]{ 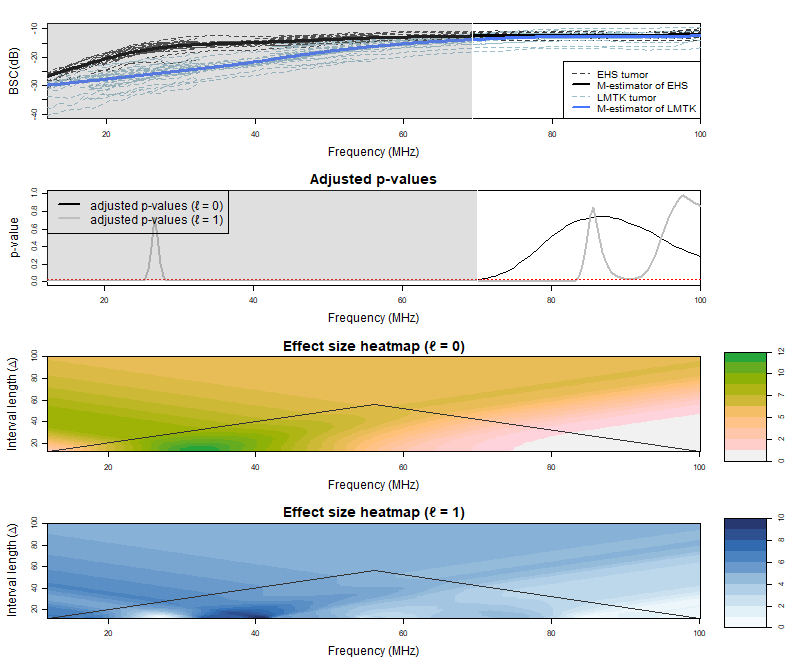}
   \caption{(Top row) The EHS and LMTK data with bold lines representing estimated functional location parameters, (second row) adjusted p-value functions $\tilde{p}_{D^0}(t)$ and $\tilde{p}_{D^1}(t)$ for testing on equality of two group parameters from trajectories of raw and the first-order derivatives, where the red dotted horizontal line represents the threshold for the level $\alpha=0.05$ test, and the gray-highlighted area displays the selected interval showing significant differences between EHS and LMTK groups. (Third and fourth rows) The robust effect size heatmap, where effect size values within the triangle region are calculated based on the symmetric choice of lower and upper extremes (integrated interval length equals to $\Delta$), while values outside the triangle region are derived from asymmetric lower and upper extremes (integrated interval length shorter than $\Delta$).}\label{fig:real_data}
\end{figure*}
We illustrate the application of the proposed robust domain selection method to the ultrasonic BSC versus frequency data for two types of mouse tumors, EHS and LMTK. The BSC data were acquired ex vivo from excised mouse tumors using single-element ultrasonic transducers with center frequencies 20, 40, and 80 MHz, leading to their corresponding coverage frequencies to be 14-33, 17-65, and 42-109 MHz, respectively. The procedure for ultrasonic scanning and BSC computation was described in \cite{Han2013}. A total of 13 EHS tumors and 13 LMTK tumors were scanned, where each tumor sample yielded three BSC versus frequency curves from 3 transducers covering different center frequencies, shown in the top panel of Figure \ref{fig:real_data}. In \cite{Park2022}, this structure was referred to as ``functional segments over predetermined subintervals in $\mathcal{T}$" and was shown to satisfy conditions of the missing at random partial sampling scheme.

The EHS and LMTK tumors represented two tumor types with distinct tumor microstructure patterns, which in theory would yield distinct BSC versus frequency functional patterns. However, the BSC functional structure may differ more significantly at some frequencies than others. It is of practical value to determine the frequency range within which the BSC curves differ most significantly between the two tumor types. 

We apply the proposed robust domain selection method for BSC trajectories collected over varying domains. After pre-smoothing the data using the kernel approach, we estimate the robust location parameter by a functional M-estimator under the robust tuning parameter set as $\delta=1$ as in simulation studies. The resulting conclusions are empirically found to be robust to the choice of $\delta$ if reasonably set between 0.8 and 1.3. Black and blue bold lines illustrated at the top panel of Figure \ref{fig:real_data} display estimated functional M-estimates from EHS and LMTK tumor groups, respectively. Seemingly, BSC behaviors at lower frequencies are relatively separable compared to those observed at the higher frequencies. To identify specific frequency ranges displaying statistically separable features, we calculated adjusted $p$-value functions by considering up to order 1 derivative as in the middle panel of Figure \ref{fig:real_data}. By applying the threshold $\alpha^*=0.025$, we could detect frequencies between 14.5 and 63.5 MHz exhibiting significant group differences. These regions are highlighted with gray in the first and second panels. We further examine effect size heatmaps, with a strong distinction around the frequencies 20 to 40 MHz under $\ell=0$ and around the frequencies 30 to 40 MHz under $\ell=1$. The frequencies above 70 MHz especially turn out to be non-separable, with their effect size close to zero under both $\ell=0$ and 1. For the practical application of BSC measurements in this example, our results recommend acquiring the data using a transducer with center frequencies around 30 MHz.

To demonstrate the practicality of the domain selection result, we investigated the classification performance between two scenarios: i) using the entire trajectories, and ii) using only trajectories over the selected domain. Table \ref{tab:data_classification} displays 5-fold cross-validation errors from LDA and QDA (implemented using FPC scores), as well as from robust probabilistic classification (RPC) \citep{Park2019}, originally developed for BSC curve classification. While BSC trajectories already exhibit relatively well-separated group behaviors with overall low classification errors, we observe further improvement when restricting to the selected domain. Even with the loss of information, focusing on regions with more separable behaviors can enhance overall classification accuracy.

\begin{table}[t!]
    \caption{5-fold cross validated classification errors based on LDA, QDA, and Robust Probabilistic Classification (RPC) for the EHS and LMTK data using trajectories over the whole and selected domain.} \label{tab:data_classification}
        \centering 
        \vspace{2mm}
    \begin{tabular}{cccc}
         \toprule
         & LDA & QDA & RPC \\
       \midrule
    Whole domain (14 - 109 MHz)   &  0.048 & 0.049   & 0.09 \\
    Selected domain (14 - 64.5 MHz)  & 0.038 & 0.037  & 0.06 \\
       \bottomrule
    \end{tabular}
\end{table}
%
%
%
\section{Conclusion}
In this paper, a robust domain selection tool is proposed for functional data containing missing segments or abnormal behaviors. It integrates the interval-wise testing approach, which controls {\textit{interval-wise}} error rate, with a functional M-estimator to calculate test statistics robust to outlying trajectories. Furthermore, we take into account multiple desirable features of trajectories in the inference process to draw the conclusions. The proposed robust effect size heat map is expected to help clinicians identify domains featuring practically meaningful separations by displaying dynamic functional group separation patterns over the scales. In the real data application, our domain selection procedure identifies frequency ranges that exhibit significant functional distinctions between EHS and LMTK tumors. We then observe improved classification accuracy when using trajectories over these selected frequencies compared to the model using data over the entire domain.

We acknowledge that deriving asymptotic inferential properties for test statistics based on penalized spline-based M-estimators is challenging. Nonetheless, simulation studies demonstrate strong performance of the proposed method across various scenarios. Future work may focus on the theoretical development to understand asymptotic inferential properties. Additionally, extending our approach to two-dimensional functional data can be another work, and for this extension, determining features that characterize two-dimensional surface behaviors would be crucial.\\
~\\

\begin{spacing}{1}
        \bibliography{ref-bib}
        \bibliographystyle{chicago}
\end{spacing}

\end{document}